\documentclass[11pt,twoside]{article}
\usepackage{graphicx,epsfig,natbib,deluxetable}
\usepackage{CS18}
\begin{document}
%
%
%
\title{X-ray Emission from Young Stars in the TW Hya Association}
%
%
\author{Alexander Brown\altaffilmark{1}, Gregory J. Herczeg\altaffilmark{2},
Thomas R. Ayres\altaffilmark{1}, Kevin France\altaffilmark{1}, \&
Joanna M. Brown\altaffilmark{1}}
\affil{$^1$Center for Astrophysics and Space Astronomy,
University of Colorado, 593 UCB, Boulder, C0 80309-0593, USA}
\affil{$^2$The Kavli Institute for Astronomy \& Astrophysics, Peking
University, Beijing 100871, China}
\begin{abstract}
%
%
The 9 Myr old TW Hya Association (TWA) is the nearest group (typical distances
of $\sim$50 pc) of pre-main-sequence (PMS) stars with ages less than 10 Myr
and contains stars with both actively accreting disks and debris disks. We
have studied the coronal X-ray emission from a group of low mass TWA common 
proper motion binaries using the {\it{Chandra}} and {\it{Swift}} satellites.
Our aim is to understand better their coronal properties and how high energy
photons affect the conditions around young stars and their role in
photo-exciting atoms, molecules and dust grains in circumstellar disks
and lower density circumstellar gas. Once planet formation is underway, this
emission influences protoplanetary evolution and the atmospheric conditions
of the newly-formed planets. 
The X-ray properties for 7 individual stars (TWA 13A, TWA 13B, TWA 9A, TWA 9B,
TWA 8A, TWA 8B, and TWA 7) and 2 combined binary systems (TWA 3AB and TWA 2AB)
have been measured. All the stars with sufficient signal require two-component
fits to their CCD-resolution X-ray spectra, typically with a dominant hot
(~2 kev (25 MK)) component and a cooler component at ~0.4 keV (4 MK). The
brighter sources all show significant X-ray variability (at a level of
50-100\% of quiescence) over the course of 5-15 ksec observations due
to flares. We present the X-ray properties for each of the stars
and find that the coronal emission is in the super-saturated rotational
domain.
\end{abstract}
%
%
\section{Introduction}

Young stars are bright X-ray and ultraviolet (UV) emitters due to strong
stellar magnetic activity fostered by rapid rotation. 
These high energy photons can greatly influence protoplanetary evolution 
and the atmospheric conditions of newly formed planets.
Stellar X-ray/EUV photons are the major ionization source over most of a
protoplanetary system \citep{alexander06}.
Observation and modeling of representative samples of young stars can 
provide important insights into protoplanet evolution, because
such data provides direct, unambiguous constraints for models.
The evolutionary trend is from gas-dominated
circumstellar physics to a solid-dominated structure where the gas component
exists as icy surfaces on planetesimals and larger bodies or in the
atmospheres of protoplanets. 

\begin{deluxetable}{ccccccccc}
\tabletypesize{\tiny}
\tablecaption{Observed TW Hya Association Members and X-ray Properties\label{table1}}
\tablewidth{0pt}
\tablehead{
\colhead{Star}  &{Spectral} & \colhead{V} &
\colhead{Binary} & \colhead{Distance} & \colhead{P$_{rot}$} &
\colhead{Exposure} & \colhead{ f$_x$ (10$^{-12}$ } & \colhead{log L$_x$} \\
 &\colhead{Type}&  & \colhead{Sep.(")} &\colhead{(pc)} &
\colhead{(days)}& \colhead{(ksec)} &
\colhead{erg cm$^{-2}$ s$^{-1}$)} & \colhead{(erg s$^{-1}$)} }
\startdata
TWA 13A [NW]&M1 V&11.5&5.1&55.6$^{+2.3}_{-2.1}$&5.56&14.57&
2.19$\pm$0.03  & 29.91$\pm$0.02 \\
TWA 13B [SE]&M1 V&12.0&\nodata &59.7$^{+2.8}_{-2.5}$&5.35 &
\nodata & 2.80$\pm$0.03  & 30.08$\pm$0.03   \\
TWA 8A [N]  &M3 V&12.2& 13 &46.9$^{+3.3}_{-2.9}$&4.65& 4.56&
3.30$\pm$0.06  & 29.94$\pm$0.04  \\
TWA 8B [S]  &M5 V&15.3&\nodata &47.1$^{+3.4}_{-3.0}$&0.78 &
\nodata  & 0.16$\pm$0.01 & 28.64$\pm$0.06  \\
TWA 9A [SE]&K6 V&11.3& 5.8&46.7$^{+6.1}_{-4.9}$&5.10& 4.56&
2.13$\pm$0.04  & 29.74$\pm$0.07 \\
TWA 9B [NW]&M3 V&14.0&\nodata &50.3$^{+6.9}_{-5.4}$&3.98 &
\nodata  & 0.19$\pm$0.01  & 28.75$\pm$0.07  \\
TWA 7      & M3 V& 11.7 & ...&34.4$^{+2.8}_{-2.2}$& 5.05 &
4.69 & 3.33$\pm$0.13  & 29.67$\pm$0.05 \\
TWA 3AB    &M4V+M4V&12.6;13.1& 1.5 &35.3$^{+2.2}_{-1.9}$ &\nodata &
 6.42 & 1.07$\pm$0.07  & 29.20$\pm$0.04    \\
TWA 2AB    & M2V+M3V&11.1& 0.6 &46.6$^{+3.0}_{-2.7}$ & 4.86&
 4.89 & 0.82$\pm$0.07 &   29.33$\pm$0.05   \\
\enddata
\end{deluxetable}

The TW Hya association (TWA) is a nearby (distances $\sim$ 50 pc) 
group of 9 Myr 
old pre-main-sequence stars that samples a crucial phase of protoplanetary 
evolution. We have studied the high energy emission from 
a group of low mass, TWA common proper motion binaries, several of which can 
be spatially resolved by {\it{Chandra}} and thus permit measurement of the 
coronal properties for the individual stars. These stars possess
extremely strong photospheric magnetic fields with typical field strengths
of 3 kG \citep{yang08} and starspots large enough to show significant
optical rotational modulation \citep{Lawson_Crause05}. These strong magnetic
fields produce a high level of coronal heating and X-ray emission and most
TWA members have been recognized initially as anomalously strong X-ray
sources \citep{webb99}.
The physical properties of the stars are listed in Table \ref{table1}.

Considerable effort has been devoted recently to obtaining better
astrometry for TWA members and this has provided vastly improved
knowledge of their distances, proper motions, and space motions.
For our sample, astrometry is provided for TWA 13 and TWA 2 by
\citet{weinberger13}, for TWA 8 by \citet{riedel14}, and for TWA 3 and TWA 7
by \citet{ducourant14}. The best astrometry source for TWA 9 is still the
Hipparcos Catalogue. The distances that we have used are listed in
Table \ref{table1}.

\section{Observations and Data Analysis}

\begin{figure}
\includegraphics[angle=90, width=5.25in]{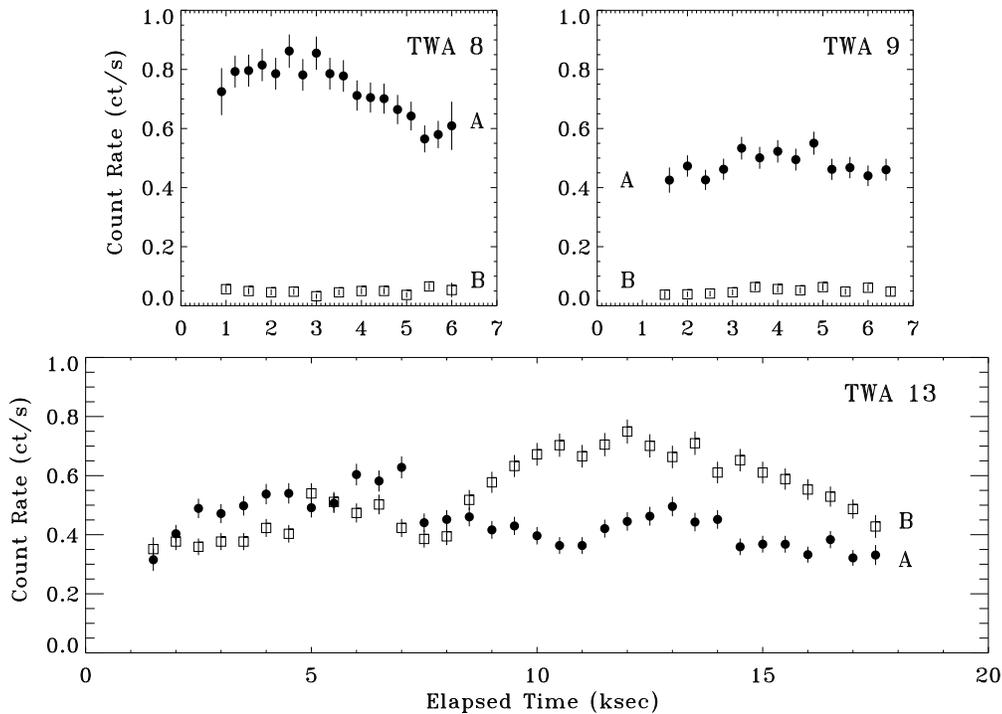}
\caption{Chandra ACIS-S3 source variability for the three common-proper-motion
binaries TWA 8AB, TWA 9AB, and TWA 13AB sampled in 500 second time-bins.
TWA 8A, TWA 13A, and TWA 13B are clearly variable.\label{fig1}}
\end{figure}

We measured the X-ray emission from nine members of the TW Hya 
association using the {\it{Chandra}} ACIS-S3 detector (Obsids: 8569, 
 8570 -- PI: Herczeg; 12389 -- PI: Brown) and the {\it{Swift}} XRT 
(Obsids: 31981001, 90207001, 90410001 -- PI:Brown), 
with the goal of a better detailed understanding of the coronal
properties of the young, low-mass (K-M) dwarf stars in the TW Hya Association.
CCD-resolution X-ray spectra with exposure times of $\sim$ 5-15 ksec
were obtained
The {\it Chandra} data were processed using CIAO Version 4.3
reduction recipes, while the {\it Swift} data were processed using
the XTOOLS data commands outlined in the {\it Swift XRT Data Reduction Guide}.
The X-ray fluxes and luminosities over the energy range 0.3-10 keV were 
measures (see Table \ref{table1}). The source variability was investigated 
when sufficient counts were available (see Fig. \ref{fig1}). 
All the resulting spectra were fitted using XSPEC Version 12.5.0
\citep{arnaud96, dorman03} and typically required
use of a two-temperature VAPEC model. At CCD-resolution the spectra are only
sensitive to changes in a few elements, particularly Fe and Ne and with a
weaker sensitivity to O. 

\section{Chandra/Swift Results}

\begin{figure}
\plotone{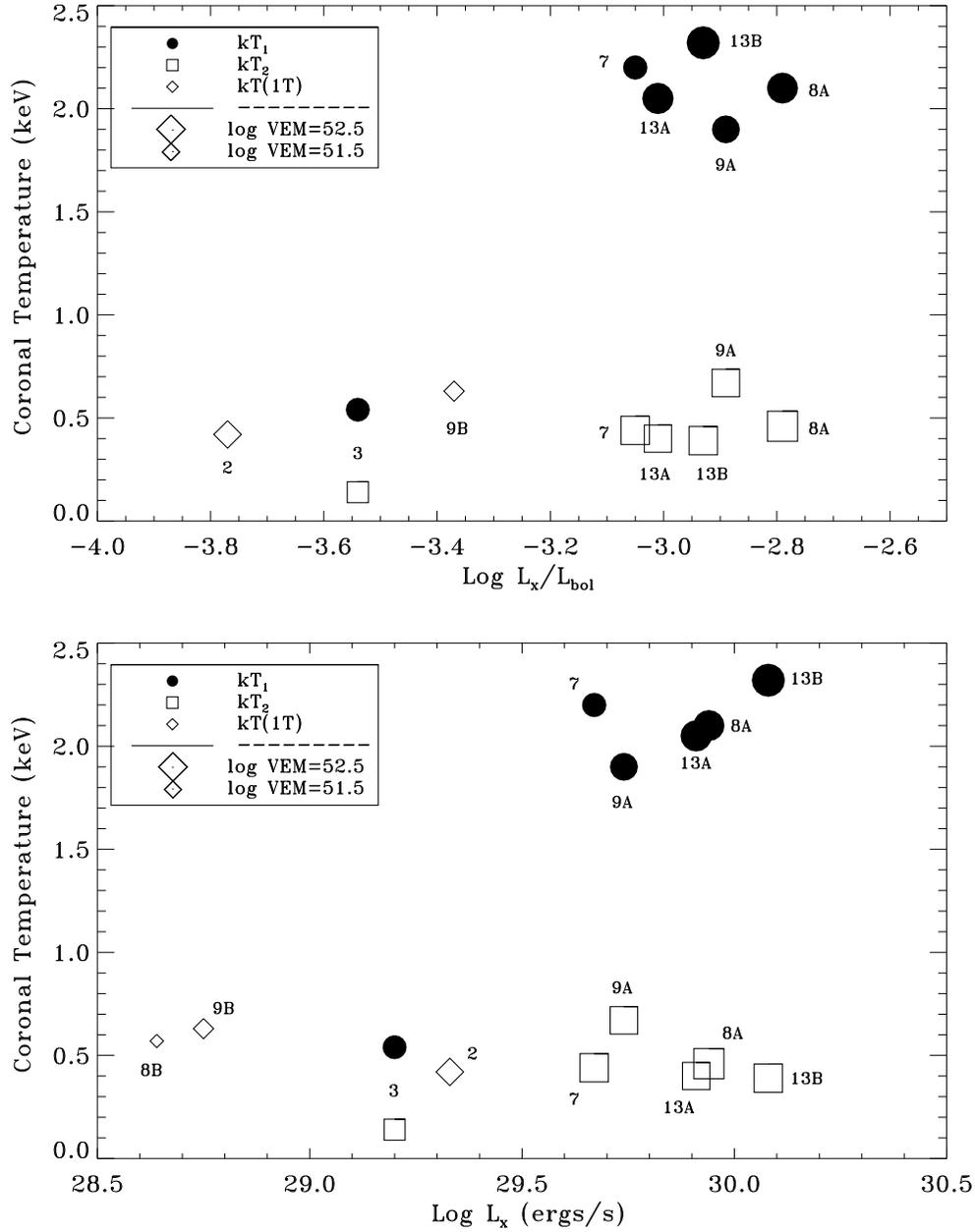}
\caption{Coronal X-ray temperature distributions 
 as a function of X-ray luminosity and
the ratio of X-ray to bolometric luminosity. Temperature components from
2-T parameterizations are shown as filled circles and open squares, with
the size of the symbol scaled to the volume emission measure. A factor
of 10 increase in VEM is shown by a doubling of the symbol size. Single
temperature parameterizations are shown by open triangles. \label{fig2}}
\end{figure}

Our basic results can be summarized as follows:

\begin{itemize}

\item[$\bullet$]{The TWA stars were all readily detected as strong 
X-ray sources. Sufficient counts are collected to determine the X-ray 
sources positions accurately and these all agree with the expected 
proper-motion-corrected optical positions. }

\item[$\bullet$]{Our {\it{Chandra}} observations resolved the TWA 13, 
TWA 8 and TWA 9 common proper motion binaries and show that the lower mass 
but more rapidly rotating secondaries TWA 8B and TWA 9B are far 
less luminous.}

\item[$\bullet$]{Stars in the TW Hya association have super-saturated coronae
where increasing rotational velocity leads to a decrease
in the X-ray luminosity (see \citet{Jeffries11} for a general discussion). 
This has significant implications for how the 
X-ray radiation fields evolves as the stars contract and spin-up.}

\item[$\bullet$]{The coronal emission is continuously varying 
(see Fig. \ref{fig1}) due to magnetic flaring, with time-resolved 
spectral fitting showing higher temperatures corresponding to higher 
count rates. Similar time-scale FUV variations are seen in 
our contemporaneous {\it{HST}} COS spectra \citep{Loyd_France14}.} 

\item[$\bullet$]{The higher luminosity (log L$_X$ = 29.5-30.1 ergs s$^{-1}$)
stars have very hot (2 keV) coronal plasma, but the less active stars only 
show a cooler (0.5 keV) coronal component, based on 2-temperature XSPEC
VAPEC spectral fitting (see Fig. \ref{fig2}). This cooler component is 
present in the spectra of all the stars.}

\end{itemize}

\acknowledgments{
This work was supported by Chandra grant GO1-12031X and NASA Swift grants
NNX09AL59G and NNX10AK88G to the University of Colorado.
}

\end{document}